\documentclass[preprint,showpacs,aps]{revtex4}
\usepackage{epsfig}

\begin{document}
\draft \title{Longitudinal excitations in quantum antiferromagnets} 
\author{Y. Xian}
% \email{yang.xian@manchester.ac.uk}
\affiliation{School of Physics and Astronomy, The University of
Manchester, Manchester M13 9PL, UK} \date{\today } 

\begin{abstract}

By extending our recently proposed magnon-density-waves to low dimensions, 
we investigate, using a microscopic many-body approach, the longitudinal 
excitations of the quasi-one-dimensional (quasi-1d) and quasi-2d Heisenberg 
antiferromagnetic systems on a bipartite lattice with a general spin 
quantum number. We obtain the full energy spectrum of the longitudinal mode
as a function of the coupling constants in the original lattice Hamiltonian
and find that it always has a non-zero energy gap if the ground state has a
long-range order and becomes gapless for the pure isotropic 1d model. 
The numerical value of the minimum gap in our approximation agrees with that 
of a longitudinal mode observed in the quasi-1d antiferromagnetic 
compound KCuF${}_3$ at low temperature. It will be interesting to compare
values of the energy spectrum at other momenta if their experimental results
are available.

\end{abstract}


\maketitle

\section{introduction}

The low temperature properties of many two-dimensional (2d) and three-dimensional 
(3d) quantum antiferromagnetic systems can be understood by Anderson's spin 
wave theory (SWT) and its extensions \cite{ande}, which 
provides correct description of the quantum corrections to the 
classical N\'eel states of the systems. For many purposes, the 
dynamics of these systems at low temperature can be considered 
as that of a dilute gas of weakly interacting spin-wave quasiparticles
(magnons) with its density corresponding to the quantum correction 
to the classical N\'eel order; also present in these systems are the 
longitudinal fluctuations consisting of the multi-magnon 
continuum \cite{hube}.

On the other hand, due to the strong quantum fluctuations, the isotropic
1d antiferromagnets with low quantum spin numbers exhibit different low temperature 
properties, such as no N\'eel-like long-range order in the ground state and 
the quite different low-lying excitation states from the 2d and 3d 
counterparts. According to the exact solutions by Bethe ansatz, the 
natural low-lying excitation states of the 1d spin-1/2  Heisenberg 
model have been shown corresponding to the spin-1/2 objects 
(spinons) which always appear in pairs, and the 
spin-wave-like excited states are actually the triplet states of 
a spinon pair \cite{cloi}, contrast to the doublet 
states from SWT. For the spin-1 Heisenberg chain, however, the triplet 
excitation states have a nonzero energy gap, first predicted by 
Haldane \cite{hald}. These theoretical predictions have been 
confirmed by the experimental results in the antiferromagnetic compound 
KCuF${}_3$ for the spin-1/2 chains \cite{tenn} and CsNiCl${}_3$ for 
the spin-1 chains \cite{buye}.

Strictly, all real systems are three dimensional when temperature 
is low enough. The antiferromagnetic compounds KCuF${}_3$ or 
CsNiCl${}_3$ are actually quasi-1d systems with very weak inter-chain 
couplings. In particular, the spin-spin couplings are ferromagnetic in 
the tetragonal basal planes of KCuF${}_3$ and 
antiferromagnetic in the hexagonal planes of 
CsNiCl${}_3$. Many parent compounds of the high-$T_c$ superconducting 
cuprates or the ion-based pnictides are also quasi-2d antiferromagnetic 
systems with very weak inter-plane couplings \cite{sham,kami}. Therefore, 
there is a 3d magnetic long-range order with a nonzero N\'eel temperature 
$T_N$ for all these systems and one expects SWT should provide a 
qualitatively correct description for some if not all the low-temperature 
dynamics of these quasi-1d or quasi-2d systems. One interesting question 
is whether or not some 1d-type excitations, such as the longitudinal part 
of the triplet spin-wave excitations of the pure 1d systems can survive in the ordered
phase when the inter-chain couplings are present. In fact, there is now ample evidence 
of the longitudinal excitation states in various quasi-1d structures with the N\'eel-like
long-range order at low temperature, including the hexagonal $ABX_3$-type 
antiferromagnets with both spin quantum number $s=1$ (CsNiCl${}_3$ and 
RbNiCl${}_3$) \cite{stei,tun} and $s=5/2$ (CsMnI${}_3$) \cite{harr,kenz} 
and the tetragonal structure of KCuF${}_3$ with $s=1/2$ \cite{lake}. 
More recently, a longitudinal mode was also observed in the dimerized
antiferromagnetic compound TlCuCl${}_3$ under pressure with a long-range N\'eel 
order \cite{rueg}. To our knowledge, no observation of any longitudinal 
mode in the quasi-2d antiferromagnets has been reported yet. Clearly, 
such longitudinal modes, which correspond to the
oscillations in the magnitude of the magnetic order parameter, are beyond the 
usual SWT which only predicts the transverse spin-wave excitations (magnons).
There have been several theoretical investigations 
in these longitudinal modes, all using the field theory approach, such 
as a simplified version of Haldane's theory for the spin-1 
systems \cite{affl} or the sine-Gordon 
theory for the spin-1/2 systems \cite{schu, essl}, and both treating the 
inter-chain couplings as perturbation. A phenomenological field theory approach 
focusing on the spin frustrations of the hexagonal lattice of the $ABX_3$-type 
antiferromagnetic systems has also been proposed \cite{plum}. Common 
to all these field theory approaches is the need to take the continuum 
limit with a number of fitting parameters. By proper choice for the values
of the fitting parameters, general agreements with the experiments mentioned
earlier have been found, although there are still some disagreements
particularly for the data away from the minimum energy gap at the 
antiferromagnetic wavevector \cite{kenz}.

We recently proposed a microscopic many-body theory for the 
longitudinal excitations of spin-$s$ quantum antiferromagnetic 
systems, using the original spin lattice Hamiltonians \cite{yx1}. 
The basic physics in our analysis is simple: by analogue to Feynmann's 
theory on the low-lying excited states of the helium-4 
superfluid \cite{feyn}, we identify the longitudinal excitation
states in a quantum antiferromagnet with a N\'eel-like order
as the collective modes of the magnon-density waves, which represent
the fluctuations in the long-range order and are 
supported by the interactions between magnons. These longitudinal 
excitation states are constructed by the $s^z$ spin operators, 
contrast to the transverse spin operators $s^\pm$ of the magnon states in Anderson's
SWT. These modes are referred to as the collective modes of the magnon-density waves 
because of the fact that $s^z$ is the magnon-density operator in these systems. 
The energy spectra of these collective modes can be easily 
derived by a formula first employed by Feynmann for the famous phonon-roton 
spectrum of the helium superfluid involving the structure factor of 
its ground state. Nevertheless, we now realize that the precise form
for the definition of the longitudinal state in our earlier work is 
not quite correct and we have now slightly modified the definition and, 
indeed, we find the corresponding values of the energy spectra in an 
approximation using the SWT ground state are in general much lower than 
before. We extend our analysis to the 1d model and find that in the 
isotropic limit the longitudinal mode has a gapless spectrum. 
Interestingly, this gapless spectrum in our approximation
is degenerate with the doublet spin-wave spectrum of SWT, hence 
making it triplet, in good agreement with the exact triplet spin-wave 
spectrum of the spin-1/2 Heisenberg model \cite{cloi}. The application 
of our analysis to the quasi-1d and quasi-2d systems is straightforward 
and hence more detailed comparison with the experiments is now possible. 
Our numerical results for the spin-1/2 quasi-1d compound KCuF${}_3$ 
show the minimum gap value of the longitudinal energy spectrum 
in agreement with the value obtained from the experiments \cite{lake}. 
This is particularly satisfactory since our analysis has no fitting 
parameters except the coupling constants in the original lattice Hamiltonian.
As our microscopic approach is able to obtain the full spectrum
of the longitudinal mode, it will be interesting to compare the values
at other regions if their experimental results are available.

We present our general theory of the magnon-density-waves briefly in Sec.~II, 
with numerical results calculated in an approximation using the SWT ground 
state for the simple cubic and square lattices and its extension to the 1d 
models in Sec.~III.  We then discuss its application to quasi-1d and 
quasi-2d systems in Sec. IV, including the quasi-1d compound KCuF${}_3$.
We summarize our work and discuss possible observations of the 
longitudinal modes in some quasi-2d 
systems in Sec.~V. We also discuss the approximations employed in our 
analysis and their possible improvements in the last section.

\section{magnon-density-waves in antiferromagnets}

We consider an antiferromagnetic system as described by an $N$-spin 
Hamiltonian $H$ on a bipartite lattice. The classical ground state 
is given by the N\'eel state with two alternating sublattices $a$ and $b$, 
where we assume the spins on the $a$-sublattice all point up in 
the $z$-direction of the spin space and the spins on the $b$-sublattice 
all point down. This N\'eel state describes the perfect two-sublattice 
long-range order. In this article, we shall exclusively use index $i$ for 
the sites of the $a$-sublattice,  index $j$ for the sites of 
the $b$-sublattice, and index $l$ for both sublattices. The excited
states are given by the spin-flipped states with respective to the
N\'eel state and are commonly referred to as magnons, the quasiparticles
of magnetic systems in general.

The quantum ground state $|\Psi_g\rangle$ of $H$ in general
differs from the classical N\'eel state by a correction, the long-range 
order is hence reduced. For many purposes, as described by the spin-wave 
theory (SWT) \cite{ande}, this quantum correction in most 2d and 3d models
is well represented by a gas of magnons whose density $\rho$ directly 
gives the correction as 
\begin{equation}
\langle s^z_i\rangle_g = s - \rho,
\end{equation}
where $s$ is the spin quantum number, $s^z_i$ is the $z$-component of spin 
operator on the lattice site $i$, and the expectation
$\langle\dots\rangle_g$ is with respect to the ground state $|\Psi_g\rangle$. 
Similarly, $\langle s^z_j\rangle_g=-s+\rho$
for the $b$-sublattice with the same density $\rho$. 
Therefore, operators $s^z$ corresponds to the magnon-density operators, 
contrast to the spin-flip operators $s^\pm$ which correspond to 
the magnon creation/destruction operators. Clearly, there are two 
types of the magnons due to the two sublattice structures.
Anderson's SWT can be most simply formulated by bosonizing the two sets of
these three spin operators, $s^z$ and $s^\pm$, on the two 
sublattices respectively. For example, the quantum correction
to the classical N\'eel state by the linear SWT gives the magnon
density of $\rho=0.078$ per lattice site for the spin-1/2 
Heisenberg model on a simple cubic lattice, and of $\rho=0.197$ per 
lattice site for the same model but on a square lattice.

Due to the interactions between the magnons, it may be
necessary to consider the states of the magnon-density waves 
(MDW). These states may not be well defined in 
the 3d systems where the magnon density is very dilute and the 
long-range order is near perfect with little fluctuations. In the 
low dimensional systems, however, the magnon density may be high 
enough to support these longitudinal waves. In terms of 
microscopic many-body theory, these MDW states are the longitudinal excitation 
states constructed by applying the density operator $s^z$ on the 
ground state in a form as $s^z|\Psi_g\rangle$, similar to Feynmann's 
theory of the phonon-roton excitation state of the helium superfluid,
where the density operator is the usual particle density operator \cite{feyn}.
These longitudinal states may be compared to the quasiparticle magnon states 
which are constructed by the transverse spin-flip operators
as $s^\pm|\Psi_g\rangle$. The above discussion underlines the main 
idea in our earlier papers \cite{yx1}, whose main purpose is to 
outline a general framework for the excitation states of both quasiparticles
and quasiparticle-density waves for a general quantum many-body
system in our variational coupled-cluster method \cite{yx2}.

In more details, following Feynmann, the MDW
excitation state with momentum $\bf q$ in an antiferromagnet is given by
\begin{equation}
|\Psi^a_q\rangle = X^a_q|\Psi_g\rangle,
\end{equation}
where excitation operator $X^a_q$, in the linear approximation, is the 
sublattice Fourier transformation of the magnon density operators $s^z_i$
of the $a$-sublattice,
\begin{equation}
X^a_q = \sqrt{\frac2N}\sum_i e^{i{\bf q\cdot r}{}_i}s^z_i, \quad q>0,
\end{equation}
with the condition $q>0$ required because of its orthogonality
to the ground state $|\Psi_g\rangle$ in which $s^z_{\rm total}=0$. 
The excitation energy spectrum in this linear approximation can be 
derived as,
\begin{equation}
E^a(q) = \frac{N^a(q)}{S^a(q)},
\end{equation}
where $N^a(q)$ is given by a double commutator,
\begin{equation}
N^a(q) = \frac12\langle [X^a_{-q},\;[H,\;X^a_q]]\rangle_g,
\end{equation}
and the state normalization integral $S^a(q)$ is in fact the structure factor of 
the $a$-sublattice,
\begin{equation}
S^a(q) = \langle X^a_{-q}X^a_q\rangle_g = 
 \frac2N\sum_{i,i'}e^{i\bf q\cdot(r_i-r_{i'})}\langle s^z_is^z_{i'}\rangle_g.
\end{equation}
Similarly, we have the MDW excitation state $X^b_q|\Psi_g\rangle$
with operator
\begin{equation}
X^b_q =  \sqrt{\frac2N}\sum_j e^{i{\bf q\cdot r}{}_j}s^z_j, \quad q>0,
\end{equation}
and the corresponding energy spectrum $E^b(q)$ for the $b$-sublattice. Due to
the lattice symmetry, the spectra $E^a_q$ and $E^b_q$ are degenerate. However,
these two excitation states are not orthogonal to each other because 
of the couplings between the spins on the $a$-sublattice and the spins on
the $b$-sublattice. We therefore need to consider their linear combinations as,
\begin{equation}
|\Psi^\pm_q\rangle = X^\pm_q|\Psi_g\rangle =
 \frac1{\sqrt2}(X^a_q\pm X^b_q)|\Psi_g\rangle,
\end{equation}
for the coupled MDW states. The corresponding energy spectra
is similarly given by $E^\pm(q)= N^\pm(q)/S^\pm(q)$ with $N^\pm(q)$ and $S^\pm(q)$
given by similar equations as Eqs.~(5) and (6) respectively using excitation
operators $X^\pm_q$ instead of $X^a_q$. It seems that we have two longitudinal
modes. But these two states $|\Psi^\pm_q\rangle$ with the energy spectra $E^\pm(q)$
are actually the same state, with one equal to another by a substitution for the
wavevector ${\bf q}\to {\bf q}+{\bf Q}$ where ${\bf Q}$ is the antiferromagnetic
wavevector of the system (e.g, ${\bf Q}=(\pi,\pi)$ for the 2d square lattice model).
This can be easily seen as the excitation operator $X^\pm_q$ are in 
fact nothing but the Fourier transformations of the (staggered) magnon density 
operators $s^z_l$ or $(-1)^ls^z_l$ respectively. We therefore only need
to consider one of them. We choose $|\Psi^-_q\rangle=|\Psi_q\rangle$ with its
energy spectrum $E^-(q)=E(q)$, and write
\begin{equation}
E(q) = \frac{N(q)}{S(q)},
\end{equation}
where $N(q)$ and $S(q)$ are calculated by using $X^-_q = X_q$ of Eq.~(8).
We notice the slight difference
between this definition of the MDW states of Eq.~(8) and that in our
earlier paper \cite{yx1} where we used the total density operator
as $\hat n_i = \frac12(2s-s^z_i+\frac1z\sum_ns^z_{i+n})$ with $z$ as the
coordination number and $n$ as the nearest-neighbor index. We now realize the use 
of operator $\hat n_i$ (or its equivalent form, $s^z_i-\frac1z\sum_ns^z_{i+n}$) is
not quite correct. Our current definition of Eq.~(8) seems more natural as 
discussed in details above. Indeed, as we will see later, the values of the 
energy spectrum of the states defined by Eq.~(8) in our approximation
are in general much lower than before, with the maximum energy values 
about half of those of the earlier results \cite{yx1}.

So far, in the above general analysis for the longitudinal excited
states, the exact ground state $|\Psi_g\rangle$ is used for the 
ground state expectation values. The only approximation comes from the 
choice of the linear form in the excitation operators of Eqs.~(3) 
and (7), and is often referred to as the single-mode approximation 
as viewed from the general expression
of the dynamic structure factor. In the case of the helium superfluid, 
the double commutator can be simply evaluated as $N(q)\propto q^2$, 
and Feynmann \cite{feyn} used the experimental results for the 
structure factor with $S(q)\propto q$ as $q\to0$ and hence derived the low-lying
phonon spectrum $E(q)\propto q$ and the gapped roton spectrum around the
peak of $S(q)$. Jackson and Feenberg, however, used the variational results 
calculated from the Jastrow-type wavefunctions and obtained
similar results \cite{jack}. In our earlier papers \cite{yx1}, 
we have demonstrated that these equations remain valid
when the exact ground state $|\Psi_g\rangle$ is replaced by a variational 
state $|\Psi_0\rangle$ and furthermore, in the case of the quantum antiferromagnets
as discussed here, our variational ground state $|\Psi_0\rangle$ by 
the so-called variational coupled-cluster method in a first order
approximation reduces to that of Anderson's 
SWT \cite{yx2}. Therefore, to this first order approximation which
is what we focus on here, we apply the SWT ground state $|\Psi^{\rm sw}\rangle$
in all of our following calculations. We like to emphasize that SWT itself
in its usual form cannot produce the longitudinal MDW excitations discussed here.
We will discuss this approximation and its possible improvement in 
the last section. We present our numerical results of the MDW spectra $E(q)$ for
several models in the following two sections: Sec.~III contains the results
for the antiferromagnetic models on the simple lattices, while Sec.~IV contains 
results for the more physical quasi-1d and quasi-2d systems.

\section{Results of magnon-density-wave spectra in simple lattices}
\subsection{The spin-$s$ $XXZ$ Heisenberg model}

In this section, we present the numerical results for the energy 
spectra of the MDW states as discussed in the earlier section
for the spin-$s$ $XXZ$ Heisenberg model on a simple cubic lattice and 
a square lattice. We then present the results for the 1d model and
discuss the convergent results in its isotropic limit.

The spin-$s$ $XXZ$ Heisenberg model on a bipartite lattice is given by
\begin{equation}
H = J\sum_{i,n}\left[\frac12(s^+_is^-_{i+n}+s^-_is^+_{i+n})+As^z_is^z_{i+n}\right],
\end{equation}
where the coupling parameter $J>0$,
index $i$ runs over all $a$-sublattice only, index $n$ runs over 
the $z$ nearest-neighbor sites, and $A$ ($\ge1$) is the anisotropy 
parameter. The usual isotropic Heisenberg model is given by $A=1$. 
The purpose of introducing the anisotropy is twofold: it is interesting 
on its own right and it also provides a way to obtain convergent
results for the 1d case in the isotropic limit as we will see later.

Using the usual spin commutation relations, it is straightforward to 
derive the following double commutator as,
\begin{equation}
N(q) = \frac12\langle[X_{-q},\;[H,\;X_q]]\rangle_g=
-\frac{zJ}2(1+\gamma_q)\langle s^+_is^-_{i+1}\rangle_g,
\end{equation}
where $\gamma_q$ is defined as usual,
\begin{equation}
\gamma_q  = \frac1z\sum_ne^{i{\rm q\cdot r}_n},
\end{equation}
with the coordination number $z$, and $\langle s^+_is^-_{i+1}\rangle_g$ is independent
of the index $i$ due to the lattice translational symmetry.
The general expression for the structure factor $S(q)$ contains an
additional cross term compared to the sublattice counterpart $S^a(q)$ as
\begin{equation}
S(q) =  \langle X_{-q}X_q\rangle_g = S^a(q)+\frac2N\sum_{i,j}
  e^{i\bf q\cdot(r_i-r_{j})}\langle s^z_is^z_{j}\rangle_g.
\end{equation}
Before we discuss any approximation, we notice that the double commutator
in general behaves as, near the antiferromagnetic wavevector $\bf Q$,
\begin{equation}
N(|{\bf q}+{\bf Q}|)\propto q^2,\quad q\to 0,
\end{equation}
similar to that of the helium superfluid \cite{feyn}.

Now we need a specific approximation for the ground state $|\Psi_g\rangle$
in order to evaluate the spin correlation functions $\langle s^+_is^-_j\rangle_g$, 
$\langle s^z_is^z_{i'}\rangle_g$, and $\langle s^z_is^z_j\rangle_g$. As 
mentioned earlier, in this article we use as our first-order approximation 
the spin-wave ground state, $|\Psi^{\rm sw}\rangle$, for these 
calculations. After defining the transverse spin correlation function $\tilde g(r)$,
\begin{equation}
\tilde g_r = \frac1{2s}\langle s^+_i s^-_{i+r}\rangle_g,
\end{equation}
we derive the following results for its Fourier transformation,
\begin{equation}
\tilde g_q = -\frac1{2}\frac{\gamma_q/A}{\sqrt{1-\gamma^2_q/A^2}},
\end{equation}
and the sublattice structure factor,
\begin{equation}
S^a(q)=\rho+\sum_{q'}\rho_{q'}\rho_{q-q'},\quad
\rho_q = \frac12\frac1{\sqrt{1-\gamma^2_q/A^2}}-\frac12
\end{equation}
with the magnon density $\rho=\sum_q\rho_q$. And, finally,
the full-lattice (staggered) structure factor is given by,
\begin{equation}
S(q) = S^a(q)+\sum_{q'}\tilde g_{q'}\tilde g_{q-q'}.
\end{equation}
We notice that in deriving the expressions of Eqs.~(17) and (18) for
the structure factors, the values for $q=0$ are excluded due to the
condition $q>0$ in the definition of $X_q$ from Eqs.~(3) and (7).
Furthermore, the integrals in the structure factor involving
function $\gamma_{q'}\gamma_{q-q'}$ clearly indicate the couplings
between magnons. In all these formulas, the summation over $q$ is given by
\begin{equation}
\sum_q = \frac1{(2\pi)^d}\int^\pi_{-\pi} d^dq,
\end{equation}
where $d$ is the dimensionality of the system. The energy spectrum $E(q)$ of Eq.~(9) is
obtained by calculating the values for $N(q)$ and $S(q)$ from the approximations
of Eqs.~(16-18). This longitudinal spectrum $E(q)$ can be compared with the 
following transverse spin-wave spectra of the linear SWT \cite{ande},
\begin{equation}
E^{\rm sw}(q) = szJA\sqrt{1-\gamma^2_q/A^2}.
\end{equation}
In the following subsections, we present numerical results using the
above approximations.

\subsection{Results for the simple cubic and square lattices}

We first consider the isotropic case $A=1$ for the simple cubic lattice 
model for which $\tilde g_1$ is calculated as
$\tilde g_1=\langle s^+_is^-_{i+1}\rangle_g\approx-0.13$.
The numerical values for $E(q)$ near $q\to0$  are similar as 
given earlier \cite{yx1}, with a large energy gap of about $0.99szJ$ at $q=0$. 
But at other values of $q$, the energies are much smaller than before due to the
different definitions of the density operator of Eq.~(8) \cite{corr}. At
the antiferromagnetic wavevector (AFWV) ${\bf Q}=(\pi,\pi,\pi)$, 
the spectrum has a larger gap of $1.40szJ$. 
As discussed before, this high energy 3d longitudinal mode may 
not be well defined and distinguishable from the multimagnon continuum.

For the square lattice model at the isotropic point $A=1$, $\tilde g_1\approx-0.28$.
Similar to the earlier results \cite{yx1}, $E(q)$ becomes gapless at both
AFWV ${\bf Q}=(\pi,\pi)$ and $q\to0$, 
due to the logarithmic behaviors from the structure factors (e.g., 
$S(q)\to-\ln q$ hence $E(q)\propto -1/\ln q$ as $q\to0$). 
However, as discussed earlier, this logarithmic gapless spectrum 
of the square lattice model in fact is quite "hard" in the 
sense that any finite-size effect, anisotropy or interplane coupling to be
discussed later, however small, will make a nonzero gap. For example, 
we consider a tiny anisotropy here with a value $A-1=1.5\times 10^{-4}$, 
which in fact is a typical value for the parent compound of the high-$T_c$
superconducting cuprate, La${}_2$CuO${}_4$
\cite{keim}, we obtain in our approximation the gap values at 
$E(Q) \sim 0.76szJ$ and $E(q) \sim 0.44szJ$ as $q\to0$, 
both much larger than the corresponding magnon gap value of $0.02szJ$ from 
Eq.~(20). We plot part of the spectrum with this anisotropy
in Fig.~1, together with the spin-wave spectrum for comparison.
The energy values at the two particular momenta $(\pi/2,\pi/2)$ 
and $(\pi,0)$ deserve attention, where $\gamma_q=0$ and the spin-wave
spectrum gives the same value of $szJ$. The longitudinal spectrum $E(q)$ at 
these two point has slightly different values, $1.36szJ$ and $1.40szJ$ respectively.
The energy difference at these two momenta has been used to indicate 
nonlinear effects due to magnon-magnon interactions in the more accurate 
calculations for the isotropic Heisenberg model \cite{zhen}. It is interesting 
to note that our longitudinal mode also show this difference.

\subsection{Results for the 1d model}

We next consider the 1d case. The SWT results in general for the isotropic 
1d case are not reliable as most integrals suffer from the well-known 
infrared divergence, e.g., the magnon density $\rho\to\infty$ as $A\to1$,
an unphysical result. Nevertheless, the value of the spin-wave spectrum of Eq.~(20)
is not far off that of the exact result by Bethe ansatz \cite{cloi}
for the spin-1/2 model despite the different degeneracies (i.e., the spin-wave
spectrum is doublet while the exact spectrum is triplet).
The infrared divergence of the spin-wave results 
also occurs for the parameter $\tilde g_1$ in the numerator of 
the energy spectra in Eq.~(9). We examine the behaviors of each 
integral in $N(q)$ and $S(q)$ in the isotropic limit $A\to1$ and find 
that they all have the similar infrared divergence. For example, by 
numerical calculations, we find that
\begin{equation}
\tilde g_1\propto -\frac1{2\pi}\ln(A-1),\quad {\rm as}\;A\to1,
\end{equation}
agrees with the analytical results using the elliptical formula \cite{welz}.
Furthermore, in the limit $q\to0$, both $S^a(q)$ and $S(q)$ behave as
\begin{equation}
S(q)\to -\frac1{2\pi}\frac{\ln(A-1)}{\sqrt{A-1}},\quad {\rm as}\;
q\to0\;{\rm and}\;A\to1.
\end{equation}
Since the divergences in the numerator $N(q)$ and the denominators $S(q)$
precisely cancel out, we obtain finite results for the energy 
spectrum $E(q)$ for the isotropic 1d model. Interestingly, we find that these
numerical values of $E(q)$ coincide precisely with those of the
linear spin-wave spectra of Eq.~(20) for all values of $q$ in the 
isotropic limit $A\to1$. Therefore, our longitudinal spectrum
and the doublet transverse spin-wave spectrum constitute a triplet, in 
good agreement with the following exact triplet spectrum for the spin-1/2 
model by Bethe ansatz first derived by des Cloizeaux and Pearson \cite{cloi},
\begin{equation}
E(q) = \frac{\pi J}2\sin q.
\end{equation}
The different factor $\pi J/2$ of the above exact result 
comparing to the value of $J$ by the linear SWT of Eq.~(20) with 
$z=2$ clearly comes from the nonlinear effects beyond our simple
approximation employed here. We also notice that our analysis here
in the approximations employed is not able to produce the Haldane 
gap for the isotropic spin-1 chain.

For the anisotropic 1d model (i.e., $A>1$), the triplet spectra
split and the values of the longitudinal spectrum $E(q)$ are larger 
than those of the doublet spin-wave spectrum, similar to the cases of 
the 2d and 3d models discussed earlier. We plot this $E(q)$ for $A=1.1$ in Fig.~2
as an example. The gaps for $E(q)$ are about $1.16szJ$ and $1.64szJ$ 
at $q=0$ and $\pi$ respectively, comparing with $0.46szJ$ of the spin-wave 
spectrum at both points.

\section{Magnon-density waves in quasi-1d and quasi-2d systems}

\subsection{Quasi-1d and quasi-2d antiferromagnets on bipartite lattices}

A generic quasi-1d and quasi-2d antiferromagnetic Hamiltonian on a bipartite 
lattice is given by,
\begin{equation}
H=J\sum_{i,n_1}{\bf s}_i\cdot {\bf s}_{i+n_1}
+J_\perp\sum_{i,n_2}{\bf s}_i\cdot {\bf s}_{i+n_2},
\end{equation}
where index $i$ as before runs over all $a$-sublattice sites with 
index $n_1$ over the nearest-neighbor sites along the chains and $n_2$ over
the nearest-neighbor sites on the basal planes, $J$ is the coupling
constant along the chains and $J_\perp$ is the counterpart on the
basal planes. We consider the model with both $J$ and $J_\perp>0$. The quasi-1d model
corresponds to the case of $J_\perp/J\ll1$, the quasi-2d model to the
case of $J_\perp/J\gg1$, and the 3d model is given by $J_\perp=J$.  
This Hamiltonian has been studied for the case of the quasi-1d systems
with $J_\perp/J\ll1$ by SWT \cite{ishi,welz}. In particular, the SWT 
ground state was used to evaluate the corrections due to the kinematic 
interactions to the order parameter $\rho$. The longitudinal modes 
were not discussed.

All of our earlier formulas for the longitudinal mode at the isotropic 
point $A=1$ remain the same after the following replacements
\begin{equation}
z \to z'=2(1+2\xi),\quad \gamma_q\to \gamma'_q =
 \frac2{z'}\left[\cos q_z+\xi(\cos q_x +\cos q_y)\right],
\end{equation}
where $\xi=J_\perp/J$. This is true also for the spin-wave spectrum 
$E^{\rm sw}(q)$ of Eq.~(20). We notice that the spin-wave spectrum is 
gapless at zone boundaries, the longitudinal mode $E(q)$ of Eq.~(9) however has
nonzero gaps for any $\xi>0$, at which there is a long-range order \cite{ishi,welz}. 
In Fig.~3, we present our results for the spectrum, denoted as $E^{\rm q1d}(q)$, 
of the quasi-1d model with $A=1$ and $\xi=1.05$ 
as an example, together with the spin-wave spectrum $E^{\rm sw}(q)$.
The gaps for $E^{\rm q1d}(q)$ at $q\to0$ and ${\bf Q}=(\pi,\pi,\pi)$ are $0.78sz'J$ and 
$1.21sz'J$ respectively. Fig.~3 also includes our results for a 
quasi-2d model with $A=1$ and $1/\xi=J/J_\perp=10^{-3}$,
denoted as $E^{\rm q2d}(q)$. The gap values
for this quasi-2d spectrum at $q\to0$ and ${\bf Q}=(\pi,\pi,\pi)$ are about
$0.47sz'J$ and $0.80sz'J$ respectively. We also notice that at the 
particular two momenta $(\pi/2,\pi/2,0)$
and $(\pi,0,0)$, where the linear spin-wave spectrum has the same
value of $sz'J$ but the longitudinal mode has slightly different values, $1.36sz'J$ 
and $1.41sz'J$ respectively, due to magnon-magnon interactions as discussed
earlier. This quasi-2d model may be relevant to the 
parent compounds of the high-$T_c$ cuprates, where the effective
interlayer couplings $J/J_\perp$ between the CuO${}_2$ planes are estimated
to be between $10^{-2}$ and $10^{-5}$ \cite{sing}.

\subsection{Quasi-1d model with KCuF${}_3$ structure}

In the experimentally well studied quasi-1d compound KCuF${}_3$, 
the strong spin couplings along the chains are antiferromagnetic 
but the weak couplings on the basal plane are ferromagnetic. 
This compound can be described by the following Hamiltonian model, 
\begin{equation}
H = J\sum_{l_a,n_1} {\bf s}_{l_a}\cdot {\bf s}_{l_a+n_1}
  -\frac {J_\perp}2\left(\sum_{l_a,n_2}{\bf s}_{l_a}\cdot {\bf s}_{l_a+n_2}
  +\sum_{l_b,n_2}{\bf s}_{l_b}\cdot {\bf s}_{l_b+n_2}\right),
\end{equation}
whose classical N\'eel state consists of two alternating planes, with all the
spins on the $a$-plane pointing up and labeled by index $l_a$ and all the spins
on the $b$-plane pointing down and labeled by index $l_b$.
In Eq.~(26), the nearest-neighbor indices $n_1$ and $n_2$ are as defined before
with $n_1$ along the chains and $n_2$ on the basal planes,
and both $J$ and $J_\perp>0$. The spin-wave spectrum is derived
as
\begin{equation}
E^{\rm sw}(q) = 2sJ\sqrt{\Delta^2_q - \cos^2 q_z},
\end{equation}
where $\Delta_q$ is defined as
\begin{equation}
\Delta_q=1+2\xi(1-\gamma_q^{\rm 2d}), \quad \xi=\frac{J_\perp}J,
\end{equation}
with $\gamma_q^{\rm 2d}=(\cos q_x+\cos q_y)/2$. It is easy to check 
that when $J=0$, we recover the spin-wave spectrum of the 2d 
ferromagnetic model and that when $J_\perp=0$, 
we recover the spin-wave spectrum of the 1d antiferromagnetic model.
For the longitudinal energy spectrum $E(q)$ of Eq.~(9), the double 
commutator is now given by a different form as
\begin{equation}
N(q) = -J(1+\cos q_z)\langle s^+_{l_a}s^-_{l_a+n_1}\rangle_g
 + 2J_\perp (1- \gamma_q^{\rm 2d})\langle s^+_{l_a}s^-_{l_a+n_2}\rangle_g,
\end{equation}
and the structure factor $S(q)$ is as given before by Eqs.~(6) and (13) 
in general forms and by Eqs.~(17) and (18) in our approximation using
the similar SWT ground state with the anisotropy parameter $A=1$. We notice that
in Eq.~(29), the two spin operators in the first correlation function
$\langle s^+_{l_a}s^-_{l_a+n_1}\rangle_g$ are from the two sublattices
respectively as before, but in the second correlation function
$\langle s^+_{l_a}s^-_{l_a+n_2}\rangle_g$, they are from the same
sublattice. So we still name the first one as before by $\tilde g_1=
\langle s^+_{l_a}s^-_{l_a+n_1}\rangle_g/2s$ but the second one
as $\tilde g'_1= \langle s^+_{l_a}s^-_{l_a+n_2}\rangle_g/2s$.
Using the similar SWT ground state, we obtain, for their
Fourier transformations,
\begin{equation}
\tilde g_q =-\frac12 \frac{\cos q_z}{\sqrt{\Delta^2_q - \cos^2q_z}}
\end{equation}
and
\begin{equation}
\tilde g'_q=\rho_q =\frac12 \frac{\Delta_q}{\sqrt{\Delta^2_q-\cos^2q_q}}-\frac12,
\end{equation}
respectively. We notice the quite different expressions for $\tilde g_q$ and 
$\tilde g'_q$ as expected. We present our numerical results for $E(q)$ in 
Fig.~4, together with $E^{\rm sw}(q)$  of Eq.~(27) for comparison, 
using the experimental values for the coupling constants, 
$J\approx 34$ meV, $J_\perp\approx 1.6$ meV and $s=1/2$ \cite{lake}. 
Different to the longitudinal modes in other systems discussed earlier, 
we find that $E(q)$ has a smaller gap of about $0.63$J$\approx21.4$ meV at 
AFWV ${\bf Q}=(0,0,\pi)$, and a larger gap of about $0.85$J$\approx28.9$ 
meV at $q\to0$. This gap value of $21.4$ meV at AFWV is about $40\%$ higher 
than the experimental value of about $15$ meV. The field theory by Essler et al. 
produces a gap value of about $17.4$ meV \cite{essl}. However, there is
uncertainty in the estimate value of the interchain coupling constant $J_\perp$.
Lake et al. seem to have used the theoretical formula Eq.~(56) in Ref.~\cite{essl}
to obtain $J_\perp=1.6$ meV $=0.047J$. By different methods \cite{sati, hira}, $J_\perp$
was estimated to be $0.01J\sim0.016J$. Using this estimate of
$\xi =J_\perp/J=0.01$, we obtain the minimum gap value of $11.9$ meV at AFWV
and $16.8$ meV at $q=0$. Naively, if we choose about the midpoint between the
values of Refs.~\cite{lake} and \cite{sati}, $J_\perp = 0.85$ meV 
with $\xi\approx0.025$, we obtain the minimum gap value of $0.49J=16.8$ meV at 
AFWV and $0.68J=23.2$ meV at $q=0$, in good agreement with the experiment for 
the minimum gap \cite{lake}. Furthermore, with this
value of $J_\perp=0.85$ meV, the linear spin-wave spectrum gap 
at ${\bf q}=(\pi,0,\pi)$ is $E^{\rm sw}(q)\approx 0.32J=10.9$ meV, very close
to the gap value of $11\pm0.5$ meV by the experiment \cite{sati}. The
longitudinal mode $E(q)$ is nearly flat in the region $(\eta,0,\pi)\sim(\eta,0,\pi)$
with $\pi\le\eta\le0$, with the gap value about $0.59J=20.1$ meV at $(\pi,0,\pi)$.
It will be very interesting indeed to compare with 
experimental results if available for the whole spectrum.

\section{summary and discussion}

In summary, we have investigated the longitudinal excitations of various quantum 
antiferromagnets based on our recently proposed magnon-density-waves. 
Our numerical results show that the
longitudinal mode always has a nonzero gap so long the system has a N\'eel-type 
long-range order and becomes gapless in in the limit of the 1d isotropic model.
In particular, the spectrum of the longitudinal mode in our approximation is 
degenerate with the doublet spin-wave spectrum of SWT in the limit of the isotropic
1d model, in agreement with the triplet spin-wave spectrum of exact results for
the spin-1/2 model by Bethe ansatz \cite{cloi}. In the 
case of the simple cubic lattice model, the 
longitudinal mode with high energy values may not be well defined since there is 
little fluctuations in the nearly perfect classical 
long-range order. In the quasi-1d and quasi-2d models, where the
quantum correction is large and the magnon density is significant, the 
magnon-density waves may be observable. Indeed, there are now ample evidence of the
longitudinal modes in several quasi-1d systems as mentioned earlier
in Sec.~I. In particular, for the quasi-1d compound KCuF${}_3$,
our value for the minimum gap is in agreement with the experimental 
value \cite{lake}. It will be interesting if more experimental results for 
the spectrum away from the minimum are available for comparison.

It is also interesting to note that the longitudinal modes were
observed in the $ABX_3$-type antiferromagnets with both $s=1$ \cite{stei,tun}
and $s=5/2$ \cite{harr,kenz}, clearly indicating that the modes are more 
general in their physics, independent of the mechanism 
which generates Haldane gap of the 1d model. The phenomenological 
field theory model with five fitting parameters employed by 
Affleck is derived from Haldane's theory of the spin-1 chain \cite{affl}. 
It will be interesting to apply our general microscopic analysis 
presented here to the $ABX_3$-type antiferromagnets 
where the basal plane is hexagonal and the corresponding N\'eel-like state
has three sublattices rather than two sublattices discussed here.
Other systems where we can apply our analysis for the magnon-density-waves
include the quasi-2d systems where the next-nearest-neighbor antiferromagnetic
couplings, in addition to the usual nearest-neighbor couplings, are present.
These additional couplings cause quantum frustrations and the N\'eel-like
order is further reduced hence greater the magnon density to
support the magnon-density waves. Of particular current interest is 
the the parent compounds of the newly discovered high-$T_c$ superconducting 
ion-based pnictides where such next-nearest-couplings are believed to be 
significant \cite{ray}.

Finally, we want to point out that there are two major approximations 
in our analysis here. The first is the linear operators $X_q$ employed 
in constructing the excitation states and the second is the SWT ground 
state $|\Psi_{\rm sw}\rangle$ employed in evaluating all the 
correlation functions involved. In regard to the first approximation, 
it is interesting to consider the case of the phonon-roton 
spectrum of the helium superfluid, where after inclusion of the nonlinear 
terms due to the couplings to the low-lying phonons (i.e., the so-called 
backflow correction), the values of the roton gap are reduced by about half 
to near the experimental values \cite{feyn,jack}. Clearly, the effects 
due to the couplings between the longitudinal modes and the gapless 
magnons in the antiferromagnetic systems also deserve further investigation.
In regard to the second approximation, i.e., the SWT ground state employ
in our calculations, improvement can be obtained by using
better ground state functions available by more sophisticated microscopic 
many-body theories such as the coupled-cluster method \cite{ray2,yx2} and,
particularly, its most recent extension where the strong correlations are included
by a Jastrow correlation factor \cite{yx3}. We believe the quasi-1d and 
quasi-2d antiferromagnetic systems as studied here are good
theoretical models from both the view point of the field theory approach 
which deal with most effectively the nonlinear effects of the 1d 
systems \cite{affl,schu,essl} and of the microscopic many-body theory approach 
which provides general, systematic techniques in dealing with many-body correlations
in plethora of quantum systems \cite{blai}. The two theoretical approaches
complement one another in study of these models and we wish to report our 
progress in these investigations in near future.

\acknowledgements
Useful discussion with P. Mitchell and Ch. R\"uegg is acknowledged.

\newpage

\begin{figure}
  \includegraphics[scale=0.5]{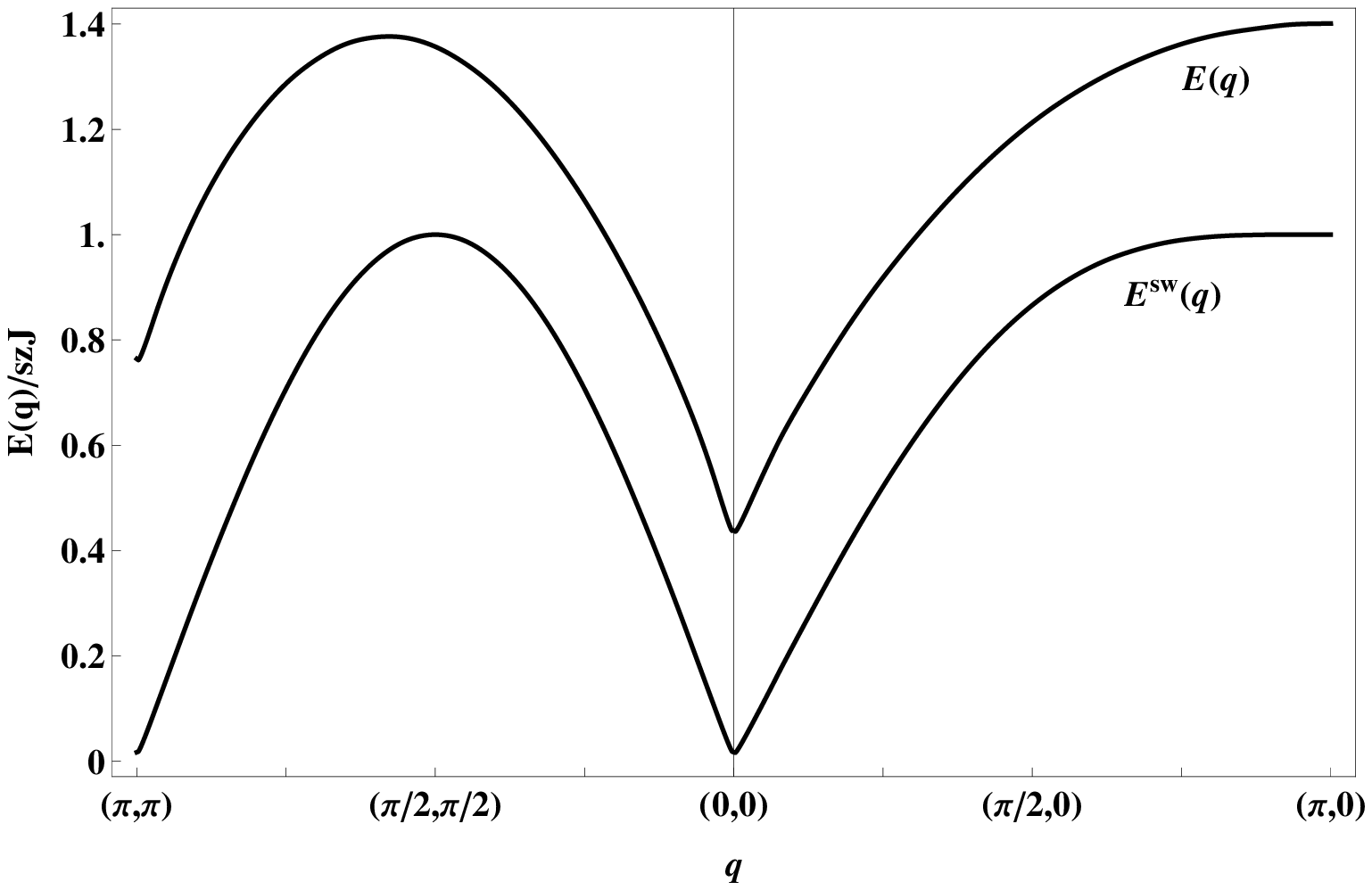}
  \vspace{2in}
   \caption{The energy spectrum $E(q)$ for the longitudinal mode of Eq.~(9) for 
the square-lattice Heisenberg model of Eq.~(10) with an anisotropy $A-1=1.5\times10^{-4}$, 
together with the linear spin-wave spectrum $E^{\rm sw}(q)$ of Eq.~(20) for 
comparison. This tiny anisotropy is a typical value for the parent compound
of the high-$T_c$ superconducting cuprate, La${}_2$CuO${}_4$.}

\end{figure}

\newpage

\begin{figure}
  \includegraphics[scale=0.5]{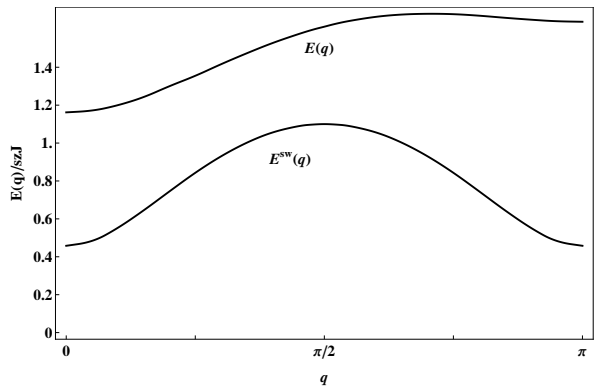}
  \vspace{2in}
   \caption{Similar to Fig.~1 but for the 1d model with the anisotropy
$A=1.1$. In the isotropic limit of $A=1$, $E(q)$ approaches to $E^{\rm sw}(q)$,
forming a triplet spectrum as described in details in text.}
\end{figure}

\newpage

\begin{figure}
  \includegraphics[scale=0.5]{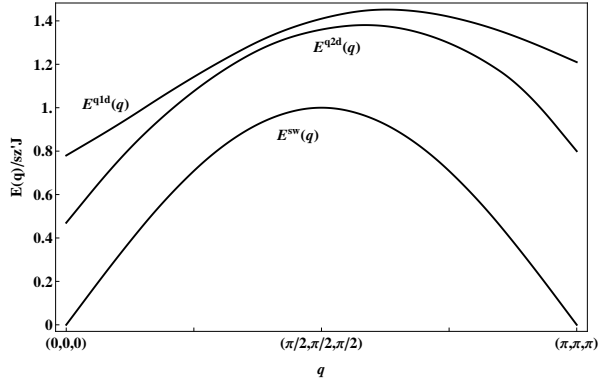}
  \vspace{2in}
   \caption{Similar to the earlier figures but for the quasi-1d and 2d systems of Eq.~(24), 
with parameter $\xi=J_\perp/J=0.01$ for the quasi-1d spectrum $E^{\rm q1d}(q)$ and
$\xi=10^3$ for the quasi-2d spectrum $E^{\rm q2d}(q)$. The spin-wave spectrum $E^{\rm sw}(q)$
is for the quasi-1d model.}
\end{figure}

\newpage

\begin{figure}
  \includegraphics[scale=0.5]{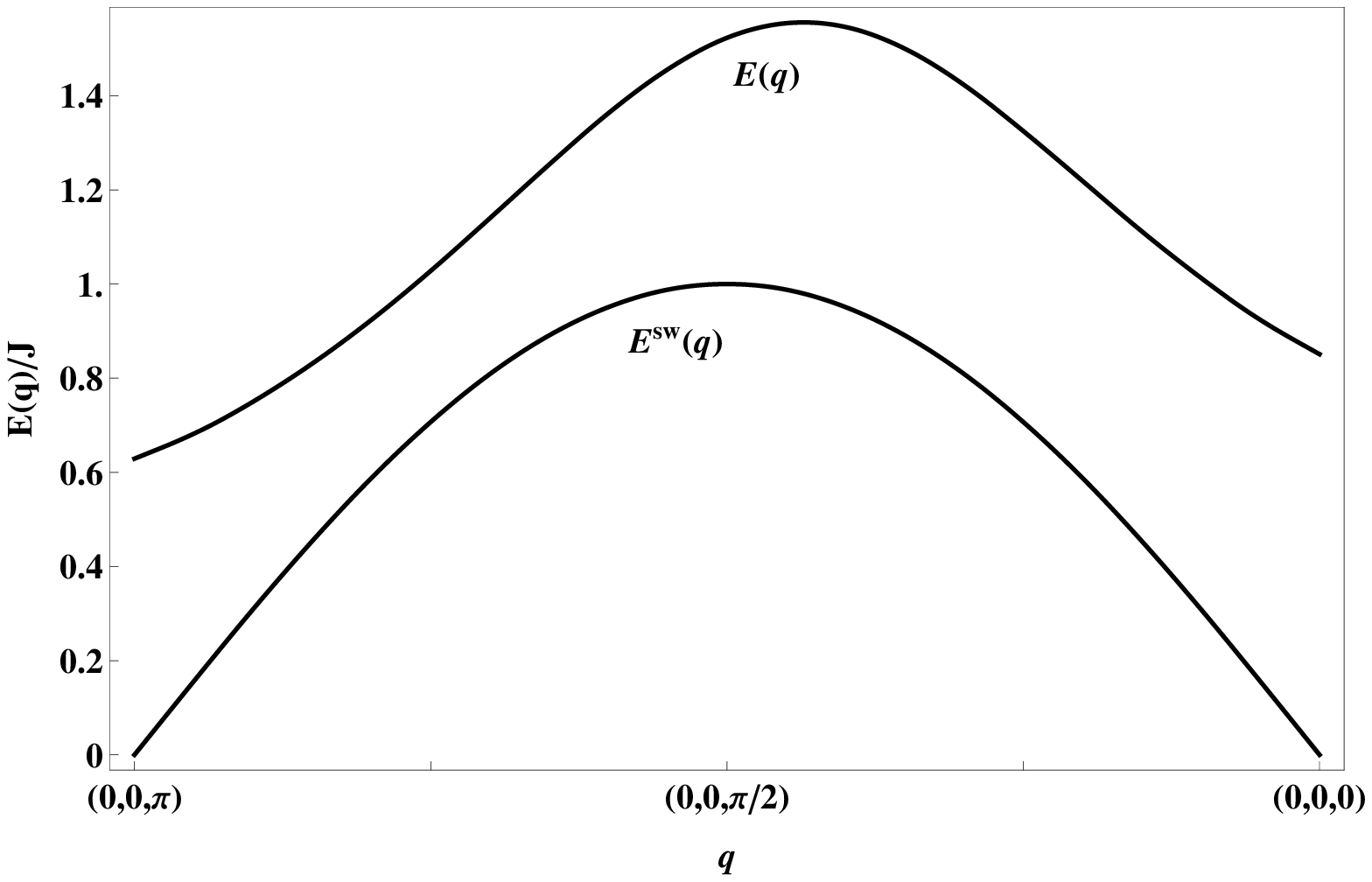}
  \vspace{2in}
   \caption{Similar to Fig.~3 but for the quasi-1d structure of KCuF${}_3$ as
described by Hamiltonian of Eq.~(26), with parameter $\xi=J_\perp/J =1.6/34$
from the experiment \cite{lake}. The spin-wave spectrum $E^{\rm sw}(q)$ is
given by Eq.~(27).}

\end{figure}


\begin{references}
\bibitem{corr} The integrals over momentum space have bounds between $(-\pi,\pi)$
as given by Eq.~(19). In Ref.~19, the bounds of $(0,\pi)$ was used by mistake
which makes no difference for most integrals but a small difference for
integrals involving $\gamma_{q'}\gamma_{q-q'}$.
\end{references}
\end{document}